\documentclass[9pt,twoside]{article}
\usepackage{amssymb}
\usepackage{latexsym}
\usepackage{epic}
\usepackage{bez123}
\usepackage{graphicx}
\usepackage{ifpdf}
\makeatletter

\@addtoreset{equation}{section} \makeatother

\newcommand{\be}{\begin{equation}}
\newcommand{\ee}{\end{equation}}

\newcommand{\R}{{\mathbb R}}

\newcommand{\flag}{strain}
\newcommand{\fl}{Str}

\newcommand{\ca}{op} 
\newcommand{\cc}{cl} 

\newcommand{\ben}{\begin{enumerate}}
\newcommand{\een}{\end{enumerate}}
\newcommand{\bit}{\begin{itemize}}
\newcommand{\eit}{\end{itemize}}
\newcommand{\edoc}{\end{document}}

\begin{document}
\date{}

\medskip


\title{{\bf\LARGE  Isocausal spacetimes may have different causal boundaries}}

\author{{\bf\large J.L. Flores$^*$,
J. Herrera$^*$,
M. S\'anchez$^\dagger$}\\
{\it\small $^*$Departamento de \'Algebra, Geometr\'{\i}a y Topolog\'{\i}a,}\\
{\it \small Facultad de Ciencias, Universidad de M\'alaga,}\\
{\it\small Campus Teatinos, 29071 M\'alaga, Spain}\\
{\it\small $^\dagger$Departamento de Geometr\'{\i}a y Topolog\'{\i}a,}\\
{\it\small Facultad de Ciencias, Universidad de Granada,}\\
{\it\small Avenida Fuentenueva s/n, 18071 Granada, Spain.}}

\maketitle
\begin{abstract} We construct an example which shows
that two isocausal spacetimes, in the sense
introduced 
recently in \cite{GS}, may have  c-boundaries which are not equal
(more precisely,  not equivalent, as no bijection between the
completions can preserve all the binary relations induced by
causality). This example also suggests that isocausality can be
useful for the understanding and computation of the c-boundary.
\end{abstract}
\begin{quote}
\end{quote}
\begin{quote}

\end{quote}

\newpage

\section{Introduction}

The {\em causal boundary}, or c-boundary for short, is a
well-known tool for the study of the conformal structure of a
spacetime and related topics such as event horizons or
singularities. The first approximation to this boundary was
introduced four decades ago by Geroch, Kronheimer and Penrose
(GKP) in the seminal paper \cite{GKP}. Since then, a long series
of redefinitions and new contributions has been carried out, and a
renewed  interest comes from the recent contributions by Harris
\cite{H1, H2, H} and Marolf and Ross \cite{MR1, MR} (see the
review in \cite{S} for   complete references). Recently, the
authors have carried out an extensive revision of both, the notion
of c-boundary and the tools for its computation
\cite{F,FHSconf,FHSst}.  So, the c-boundary can be regarded now as
a useful and consistent notion, which is well related to other
geometric objects.  Along this paper we understand by c-boundary
the last redefinition in \cite{FHSconf}. Nevertheless, the
properties to be considered here appear at a much more basic level
(say, whenever Harris' universal properties for the partial
boundaries are satisfied \cite{H1}). So, they are valid for any
definition of the c-boundary obtained by using the basic
ingredients in the seminal GKP construction and, in particular,
for all the previous redefinitions of the c-boundary along the
literature.

\smallskip
\indent Some years ago, Garc\'{\i}a-Parrado and Senovilla
\cite{GS} introduced the notions of {\em causal mapping, causal
relation} and {\em isocausality} for two spacetimes $V, V'$.
Namely, $V$ is {\em causally related} to $V'$, denoted $V\prec
V'$, if there exists a diffeomorphism $\Phi: V\rightarrow V'$
which is a {\em causal mapping}, that is,  such that all the
future-directed causal vectors of $V$ are mapped by the
differential of $\Phi$ into future-directed causal ones of $V'$.
Then, $V$ is {\em isocausal} to $V'$ if $V\prec V'$ and $V'\prec
V$. In that article and subsequent developments \cite{GSa, GSb,
GSan}, many applications and properties of such notions were
carried out. Recall that isocausality is a generalization of
conformal equivalence, adding more flexibility. This flexibility
yields appealing properties, as the fact that any spacetime is
{\em locally isocausal} to Lorentz-Minkowski one, even if it is
not conformally flat. So, isocausality preserves some relevant
global properties associated to the conformal structure, but not
all of them ---as stressed in \cite{GSan} for the case of two
levels of the causal ladder of spacetimes.

\smallskip
\indent It was also suggested          { in \cite[Sect. 6]{GS} }
that causal mappings could be used to obtain {\em causal
extensions} and boundaries for spacetimes as a generalization of
the (Penrose) {\em conformal boundary}. { Concretely, a {\em
causal extension} is an embedding of the spacetime in a larger one
such that the former is isocausal to its image in the larger.
Clearly, a boundary can be then  naturally associated to such an
extension (here, we will avoid the name {\em causal boundary} for
this last boundary, in order to avoid confusions with the
c-boundary).} Recall that, in spite of its generalized usage in
General Relativity, the conformal boundary has serious problems of
existence and uniqueness. The problems come from the fact that, in
order to find a reasonable conformal boundary, one has to find an
appropriate open conformal embedding of the spacetime in some
({\em aphysical}) spacetime. It is not clear when such an
embedding will exist and, in this case, if the properties of the
corresponding boundary will be independent of the
embedding\footnote{{ An example of the difficulties can be found
in the  recent article \cite{Chr}. In order to ensure uniqueness,
some technical assumptions (which involve any pair of lightlike
curves) must be assumed. Remarkably,  the existence of a maximal
conformal extension is also ensured in \cite{Chr}. However, this
does not exclude the possibility that no extension exists, nor
ensures a priori  good properties for such an extension.}}. The
flexibility of causal mappings and isocausal properties, allows to
check their existence much more easily than their conformal
counterparts, even though with a cost of uniqueness.

\smallskip
\indent In the present note, we explore the connections between
the c-boundary and the notion of isocausality by means of a
concrete example. This example firstly shows that {\em two
isocausal spacetimes may have different c-boundaries}. That is,
even though the c-boundary relies on the global conformal
structure of the spacetime, it is not an object naturally
invariant by isocausality.  At a first glance, this property would
seem a  drawback for the notion of isocausality. On one hand,
isocausality  would be insufficient to distinguish between two
spacetimes with different asymptotic causal behaviors. On the
other, the boundaries obtained by using different causal
extensions appear as extremely non-unique ---as the causal
extensions of isocausal spacetimes with different c-boundaries,
may  look very different. However, a deeper study suggests that,
when the causal extensions are compared with the conformal ones,
these properties are not a disadvantage. Notice that, essentially,
the conformal boundary becomes interesting when it agrees with
some intrinsic element of the spacetime, and conditions to ensure
this agreement are commonly imposed (see \cite{AH,FHSconf}). But
the most important intrinsic element of the spacetime which may
match with the conformal boundary is  the c-boundary; so,
basically, the conformal boundary becomes useful
 as an auxiliary tool to compute the more general c-boundary.
On the contrary, the properties which remain true for all the
elements of a class of isocausal spacetimes (in particular, the
possible similarities  of their c-boundaries or of the boundaries
obtained through causal extensions), become a genuinely new type
of information, which reveals new connections among
non-conformally related spacetimes. In fact, a closer look at our
example in this article, suggests that {\em causal mappings and
isocausality may yield a very valuable information on the
c-boundary}. Here, we explain this possibility only for our
particular example, in order to provide a natural intuitive
picture. By using the machinery introduced in \cite{FHSst}, this
idea will be developed technically in a further work by the authors.

\section{The example}

Typical background and terminology in Lorentzian Geometry as in
\cite{BEE, MS, O} and on causal boundaries as in
\cite{BEE,FHSconf, GS2} will be used. From the technical
viewpoint, our example will be very simple, and the c-boundary
will be handled at a very elementary level. Basically, the idea to
construct the c-boundary $\partial V$ of a (strongly causal)
spacetime $V$ starts by defining its {\em future causal boundary}
$\hat\partial V$ and the dual past one $\check\partial V$. The
former is the set of all the {\em TIP}'s (terminal indecomposable
past subsets) of $V$, where any TIP can be regarded as the
chronological past $I^-[\gamma ]$ of some future-directed
inextensible timelike curve $\gamma$ (an obvious dual definition
appears for the elements of $\check\partial V$, or {\em TIF}'s). A
long-standing problem for the definition of the c-boundary appears
when one realizes that, eventually, some points in $\hat\partial
V$ must be paired with some others in $\check\partial V$. Even
though this problem can be solved satisfactorily \cite{FHSconf},
we will not worry about it. In fact, our example will be robust,
in the  sense that even the partial boundary $\hat\partial V$ will
not be preserved by isocausality. Moreover, our concrete example
is bidimensional, and the TIPs can be also  generated as the
chronological past of (piecewise smooth) lightlike curves, instead
of timelike ones ---this will be straightforward here, however,
one can find in \cite[Proposition 2]{JHEP} and \cite[Sect.
3.5]{FHSconf} a precise justification. So, the picture of the
c-boundary is simplified, as in dimension two the (smooth)
lightlike curves must lie in two families of geodesics.

\smallskip
\noindent { {\em 2.1 Abstract properties}}. Our aim is to endow
the manifold $V=\R\times (-\infty,0)$ with three metrics
$g_{\cc}$, $g$, $g_{\ca}$ satisfying the following properties:
\begin{itemize}

\item[(i)] $g_{\cc}\prec_0 g\prec_0 g_{\ca}$ where the symbol
$\prec_0$ means that the future causal cones of the metric at the
left-hand side are included in the ones of the metric at the
right-hand  one (i.e., the identity in $V$ is a causal mapping
from $V$ endowed with the left metric to $V$ endowed with the
right one).

\item[(ii)] $g_{\cc}$ and $g_{\ca}$ are  conformally related. So
the future causal boundaries $\hat{\partial}_{\cc} V,
\hat{\partial}_{\ca} V$ for, resp., $V_{\cc}:=(V,g_{\cc})$ and
$V_{\ca}:=(V,g_{\ca})$ agree and, taking into account property
(i), $(V,g)$ is isocausal to $V_{\cc}$ (and $V_{\ca}$). Moreover
$g_{\cc}$ and $g_{\ca}$ will be simple standard static metrics, so
that its causal boundary will be easily computable.

\item[(iii)] $g$ presents a future causal boundary $\hat
\partial V$  ``strictly greater'' than the one of $g_{\cc}$ (or
$g_{\ca}$), in a precise sense explained below. { Essentially, a
segment of {\em causally but not chronologically} related points,
appears for $\hat
\partial V$ where only a point (in a timelike part of the boundary) existed for
$\hat{\partial}_{\cc} V$ and $ \hat{\partial}_{\ca} V$.}
\end{itemize}
Note that the non-preservation of the c-boundary by isocausality
follows from these properties. So, once the metrics are achieved,
we will pass to discuss the interplay between the c-boundary and
isocausality.

\smallskip
\smallskip

\noindent { {\em 2.2 Explicit construction}}. Define the metrics
$g_{\cc}$, $g$, $g_{\ca}$ on $V=\R\times (-\infty,0)$ in the
following way:
\[
g_{\cc}=-dt^2+dx^2,\qquad g=-dt^{2}+\beta(t/x)dx^{2},\qquad
g_{\ca}=-dt^2+(1/4)\;dx^2,
\]
where $\beta:\R\rightarrow (0,\infty)$ is a smooth function which
satisfies:

\begin{itemize}\item  $\beta(u)\equiv 1/4$ if $u(=t/x)\leq 1/2$, that
is, $g=g_{\ca}$ in the region $x\leq 2t $.

\item  $\beta(u)\equiv 1$ if $u\geq 1$, that is, $g=g_{\cc}$ in
the region $t \leq x (< 0)$.

\item $\beta$ increases strictly from $1/4$ to $1$ on the interval
$ 1/2\leq u\leq 1$, so that the causal cones of $g_{\cc}$ (resp.
of $g$) are strictly contained in the ones of $g$ (resp. of
$g_{\ca}$) in the region $2x<2t<x$.
\end{itemize}
Note that the announced property (i) becomes clear from the
properties of $\beta$. About (ii), the conformal relation between
$g_{\cc}$ and $g_{\ca}$ is also obvious. Moreover, the future
causal boundary $\hat
\partial_{\cc}V$  can be represented by  two  lines ${\cal T}, {\cal J}^+$ with a common endpoint
$i^+$, which is the TIP equal to all $V$ (see \cite{H,AF,FH,FHSst}
for much more general computations, which include the  c-boundary
of all the standard static spacetimes). More precisely, the TIPs
which constitute ${\cal T}$ are the chronological past of all the
future-directed lightlike geodesics $\rho$ with endpoint at $x=0$.
${\cal T}$ is timelike in the sense that any two distinct TIPs $P, P' \in
{\cal T}$ satisfy either $P \overline{\ll} P'$ or $P' \overline{\ll} P$,
where the {\em extended chronological relation} $\overline{\ll}$
can be defined here as: $P \overline{\ll} P'$ if and only if there
exists some $p'\in P'$ such that $p\ll p'$ for all $p\in P$. It is
also clear that, for the {\em (future) chronological topology} on
$\hat \partial V$ (which here reduces to the point set convergence
of the corresponding TIP's as subsets of $V$, see \cite{FH,
FHSconf, FHSst})
${\cal T}$ will be homeomorphic to $\R$. That is, in the
following, ${\cal T}$ will be identified with  $\R\times \{0\}$
(each $P\in {\cal T}$ is identified with the endpoint in $\R
\times \{0\}$ of the lightlike geodesic whose past is equal to
$P$), and this identification holds at the point set,
chronological and topological levels. The TIPs which constitute
${\cal J}^+$ are the chronological pasts of all the
future-directed lightlike $\rho$ as above which goes to infinity
(reaching arbitrarily large values of $-x$). We will not pay
attention to this line ${\cal J}^+$, but we point out that it is
{\em horismotic}. This means that any two distinct TIPs $P, P'\in
{\cal J}^+$ are {\em horismotically related}, i.e. they satisfy
either $P\subset P'$ or $P'\subset P$, but neither $P
\overline{\ll} P'$ nor $P' \overline{\ll} P$.

\smallskip

\indent For the property (iii), let us focus on the timelike line
${\cal T}$, identified with $\R\times \{0\}$. Our aim is  to prove that,
in addition to this timelike line, the future causal boundary
$\hat\partial V$ of $(V,g)$ contains other boundary points
$P=I^-[\rho]$ such that $(0,0)$ is the endpoint of the generating
future-directed lightlike curve $\rho$.

\smallskip

\begin{figure}
\centering
\ifpdf
  \setlength{\unitlength}{1bp}%
  \begin{picture}(197.00, 241.73)(0,0)
  \put(0,0){\includegraphics{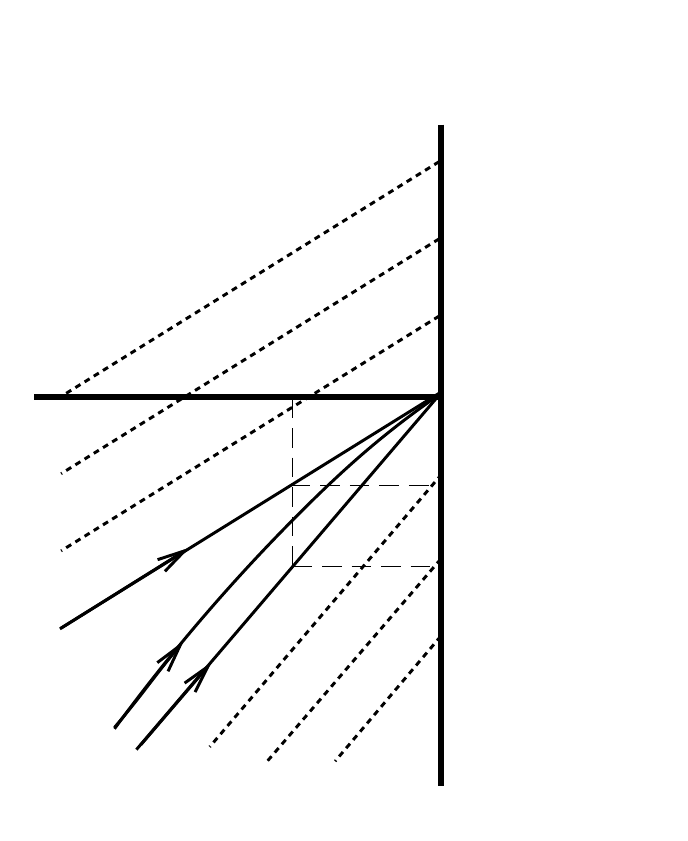}}
  \put(10.94,75.68){\fontsize{9.39}{11.27}\selectfont $\gamma_{-1/2}$}
  \put(38.24,18.76){\fontsize{9.39}{11.27}\selectfont $\gamma_{-1}$}
  \put(29.08,44.82){\fontsize{9.39}{11.27}\selectfont $\gamma_t$}
  \put(68.16,223.04){\fontsize{16.66}{20.00}\selectfont $V$}
  \put(133.44,193.75){\fontsize{10.41}{12.50}\selectfont $t$}
  \put(5.67,130.24){\fontsize{10.41}{12.50}\selectfont $x$}
  \put(82.74,7.51){\fontsize{8.54}{10.24}\selectfont $g_{\cc}$}
  \put(35.31,179.71){\fontsize{8.54}{10.24}\selectfont $g_{\ca}$}
  \put(128.20,99.93){\fontsize{10.41}{12.50}\selectfont $-\frac{1}{2}$}
  \put(129.31,76.59){\fontsize{10.41}{12.50}\selectfont $-1$}
  \put(80.40,130.83){\fontsize{10.41}{12.50}\selectfont $-1$}
  \end{picture}%
\else
  \setlength{\unitlength}{1bp}%
  \begin{picture}(197.00, 241.73)(0,0)
  \put(0,0){\includegraphics{isocausalidad1}}
  \put(10.94,75.68){\fontsize{9.39}{11.27}\selectfont $\gamma_{-1/2}$}
  \put(38.24,18.76){\fontsize{9.39}{11.27}\selectfont $\gamma_{-1}$}
  \put(29.08,44.82){\fontsize{9.39}{11.27}\selectfont $\gamma_t$}
  \put(68.16,223.04){\fontsize{16.66}{20.00}\selectfont $V$}
  \put(133.44,193.75){\fontsize{10.41}{12.50}\selectfont $t$}
  \put(5.67,130.24){\fontsize{10.41}{12.50}\selectfont $x$}
  \put(82.74,7.51){\fontsize{8.54}{10.24}\selectfont $g_{\cc}$}
  \put(35.31,179.71){\fontsize{8.54}{10.24}\selectfont $g_{\ca}$}
  \put(128.20,99.93){\fontsize{10.41}{12.50}\selectfont $-\frac{1}{2}$}
  \put(129.31,76.59){\fontsize{10.41}{12.50}\selectfont $-1$}
  \put(80.40,130.83){\fontsize{10.41}{12.50}\selectfont $-1$}
  \end{picture}%
\fi \caption{\label{fig1}Computed with the metric $g$, the curves
$\gamma_{-1/2}, \gamma_{t}$ and $\gamma_{-1}$ are lightlike and
define different TIPs. These TIPs are naturally associated to the
point $(0,0)$. However, the point $(0,0)$ is associated only to
one TIP when the metric $g_{\ca}$ or $g_{\cc}$ is considered
(recall that these two metrics are conformal).}
\end{figure}

\indent  Consider the lightlike vector field
$X(t,x)=(\sqrt{\beta(t/x)},1)$ for $g$. All the integral curves of
$X$ can be written as $\gamma_t(s)=(r_t(s),s)$, with $s<0$ and
$r_t: (-\infty, 0)\rightarrow\R$ satisfying:
\begin{equation}\label{P}
\left\{
\begin{array}{l}
\dot{r}_{t}(s)=\sqrt{\beta\left(\frac{r_{t}(s)}{s}\right)} \\
r_{t}(-1)=t
\end{array}
\right.
\end{equation}
(see Figure \ref{fig1}). Note the following properties of the curves
$\gamma_t$:
\begin{itemize}\item[(a)] For
$t_1<t_2$, necessarily $r_{t_1}(s)<r_{t_2}(s)$, as
$r_{t_{1}}(-1)=t_{1}<t_{2}=r_{t_{2}}(-1)$ and $\gamma_{t_1}$ does
not intersect $\gamma_{t_2}$. \item[(b)]
$\gamma_{-1/2}(s)=(s/2,s)$ and $\gamma_{-1}(s)=(s,s)$ for all
$s<0$, and thus, any  intermediate $\gamma_t$ satisfies:
$$\lim_{s\rightarrow 0}\gamma_t(s)=(0,0) \quad \quad \forall t\in
[-1,-1/2]. $$ \item[(c)] $I^-[\gamma_{t_1}]\subsetneq
I^-[\gamma_{t_2}]$ for all $t_{1}<t_{2}$. 
\end{itemize}
In fact, (a) and (b) are direct consequences of the definition of
$\gamma_t$. The property (c) is a consequence of (a) and the
following characterization: \be\label{ei}
I^-[\gamma_t]=\{(t',s):t'<r_t(s)\} \quad \quad \forall t\in \R.
\ee The inclusion $\supset$ follows because, for the metric $g$,
$t'<r_t(s)$ implies $(t',s)\ll (r_t(s),s)$. For $\subset$, recall
that $V\backslash\{\gamma_t(s): s<0\}$ has two connected
components, and the right-hand side of (\ref{ei}) is equal to one
of them. Any past-directed timelike curve $\alpha$ starting at a
point $p$ on $\gamma_t$ must enter initially in this region (as
any tangent vector in the past timelike cone at $p$, points to
it). Moreover, $\alpha$ cannot touch $\gamma_t$ at a distinct
(first) point $q$, as $\alpha$ and $\gamma_t$ would intersect
transversally and, so, the velocity $\alpha'$ would point out to
the future on $q$. As consequence $\alpha$ remains totally contained in that region up to the initial point $p$.

\smallskip

\begin{figure}
\centering
\ifpdf
  \setlength{\unitlength}{1bp}%
  \begin{picture}(331.24, 227.12)(0,0)
  \put(0,0){\includegraphics{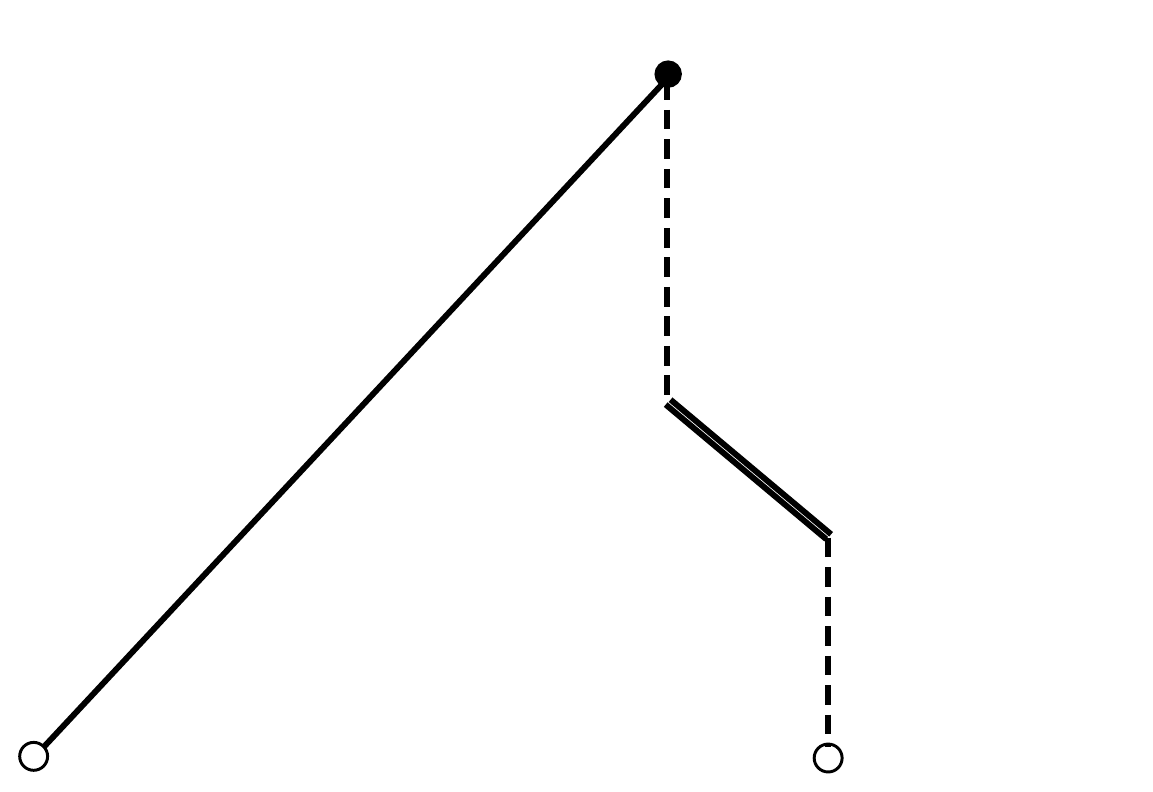}}
  \put(183.32,213.10){\fontsize{10.69}{12.83}\selectfont $i^+$}
  \put(260.06,121.26){\fontsize{12.83}{15.39}\selectfont ${\cal T}_{\fl}$}
  \put(85.71,148.65){\fontsize{12.83}{15.39}\selectfont ${\cal J}^+$}
  \put(196.39,78.59){\fontsize{10.69}{12.83}\selectfont {\fl}}
  \end{picture}%
\else
  \setlength{\unitlength}{1bp}%
  \begin{picture}(331.24, 227.12)(0,0)
  \put(0,0){\includegraphics{bordecausalfuturo}}
  \put(183.32,213.10){\fontsize{10.69}{12.83}\selectfont $i^+$}
  \put(260.06,121.26){\fontsize{12.83}{15.39}\selectfont ${\cal T}_{\fl}$}
  \put(85.71,148.65){\fontsize{12.83}{15.39}\selectfont ${\cal J}^+$}
  \put(196.39,78.59){\fontsize{10.69}{12.83}\selectfont {\fl}}
  \end{picture}%
\fi \caption{\label{fig2}Structure of the future causal boundary
for $(V,g)$. The part of the boundary ${\cal T}_{\fl}$ is composed
by two timelike lines and a lightlike one, denoted by {\fl} in the
picture, which corresponds with the {\em \flag} {\fl}$=\{P_t:\;-1\leq
t\leq -1/2\}$.}\end{figure}

\indent From the properties (b) and (c), different TIPs
$P_t:=I^{-}[\gamma_{t}]$, with  $t\in [-1 ,-1/2]$, become
naturally associated to the point $(0,0)$ (which was identified
with a point of $\hat\partial_{\cc}V$). This implies the required
property (iii). In fact, the description of the the boundary
$\hat{\partial} V$ for $g$ is similar to the one of
$\hat{\partial}_{\cc}V$. However, now in the analog to the
timelike line ${\cal T} \subset \partial_{\cc}V$, the boundary
point associated to $(0,0)$ must be replaced by all the TIPs in
the {\em \flag} \fl $:= \{P_t: -1\leq t\leq -1/2\}$. So, we can
regard ${\cal T}_{{\rm \fl}}= \left((\R\backslash
\{0\})\times\{0\}\right) \cup $ {\fl}, as a part of
$\hat{\partial} V$ (see Fig. \ref{fig2}). Recall that all the
points in the {\flag} are horismotically related. So, ${\cal T}_{{\rm
\fl}}$ differs from ${\cal T}$ from the chronological viewpoint
(there exists no bijection from ${\cal T}_{\rm \fl}$ in ${\cal T}$
which preserves the chronologically and horismotically related
points). Nevertheless, if one replaces the whole {\flag} by any
of its elements, this bijection appears naturally. Summing up, the
claimed property (iii), as well as the non-equivalence of
$\hat{\partial} V$ and $\hat{\partial}_{\cc}V$, are justified in a
precise way.

\smallskip
\smallskip

\noindent { {\em 2.3. Final discussion}}. We can understand the
behavior of the causal boundary in the previous example as
follows.

\smallskip

\indent Consider two causally related spacetimes $V_1\prec_0 V_2$
(we will write $I_1^-, I_2^-$ instead of $I^-$ in each spacetime).
A natural map between the future boundaries $\hat j: \hat\partial
V_1\rightarrow \hat\partial V_2$ can be defined by taking into
account that if $P\in \hat\partial V_1$ then $I_2^-(P)\in
\hat\partial V_2$. In fact, if $P=I_1^-[\gamma]$ for some
inextendible future-directed timelike curve $\gamma$, then
$\gamma$ must be timelike also for $V_2$, and
$I_2^-[\gamma]=I_2^-(P)$. So, we can define:
$$\hat j(P):=I_2^-(P), \quad \qquad \forall P\in \hat\partial V_1.$$
Nevertheless, $\hat j$ may be very bad-behaved, even if $V_1$ and
$V_2$ are isocausal. Concretely, our example above shows that the
map $\hat j_{\cc}: \hat\partial V_{\cc} \rightarrow \hat{\partial}
V$ associated to $V_{\cc} \prec_0 V$ cannot be continuous (nor
surjective), as it induces a map ${\cal T}\rightarrow {\cal
T}_{{\rm \fl}}$ where  $\hat j(0,0)$ chooses just the point
$I^{-}(\gamma_{-1})$ of the {\flag}. Even more, the map
$\hat j_{\ca}: \hat\partial V \rightarrow \hat{\partial}_{\ca}V$
associated to $V \prec_0 V_{\ca}$ is continuous, but it is not
injective, as all the {\flag} is mapped into $(0,0)$. It is
worth pointing out that, in spite of these properties, the
composition  $\hat j_{\ca}\circ \hat j_{\cc}: \hat\partial_{\cc} V
\rightarrow \hat\partial_{\ca} V$ is an isomorphism (a
homeomorphism which also preserves the chronological relation).
Our example shows that,  this nice last property does not imply a
straightforward good relation between $\hat{\partial} V$ and
$\hat{\partial}_{\cc}V$.

\smallskip

\indent However, the example suggests another possibility. Assume
that all the elements in the {\flag} of $\hat\partial V$ were
identified to a single one. Then $\hat{\partial}_{\cc}V$ would be
naturally embedded in this quotient space (in this particular
example, they would be naturally isomorphic). In this sense, the
boundary $\hat{\partial}_{\cc}V$ yields an important information
about the boundary $\hat\partial V$, namely:
$\hat{\partial}_{\cc}V$ represents the quotient of a part of
$\hat\partial V$ (alternatively, $\hat\partial V$ can be seen as
an enlargement of $\hat{\partial}_{\cc}V$). At what extent is this
property generalizable? We will prove  that it can be extended to
a wide family of spacetimes which are isocausal to the standard
stationary ones. However, the computation of such boundaries
requires the machinery on Finsler metrics and Busemann functions
developed in \cite{FHSst}. So, it is postponed to a forthcoming
paper.

\section*{Appendix}

\begin{figure}
\centering
\ifpdf
  \setlength{\unitlength}{1bp}%
  \begin{picture}(443.32, 256.66)(0,0)
  \put(0,0){\includegraphics{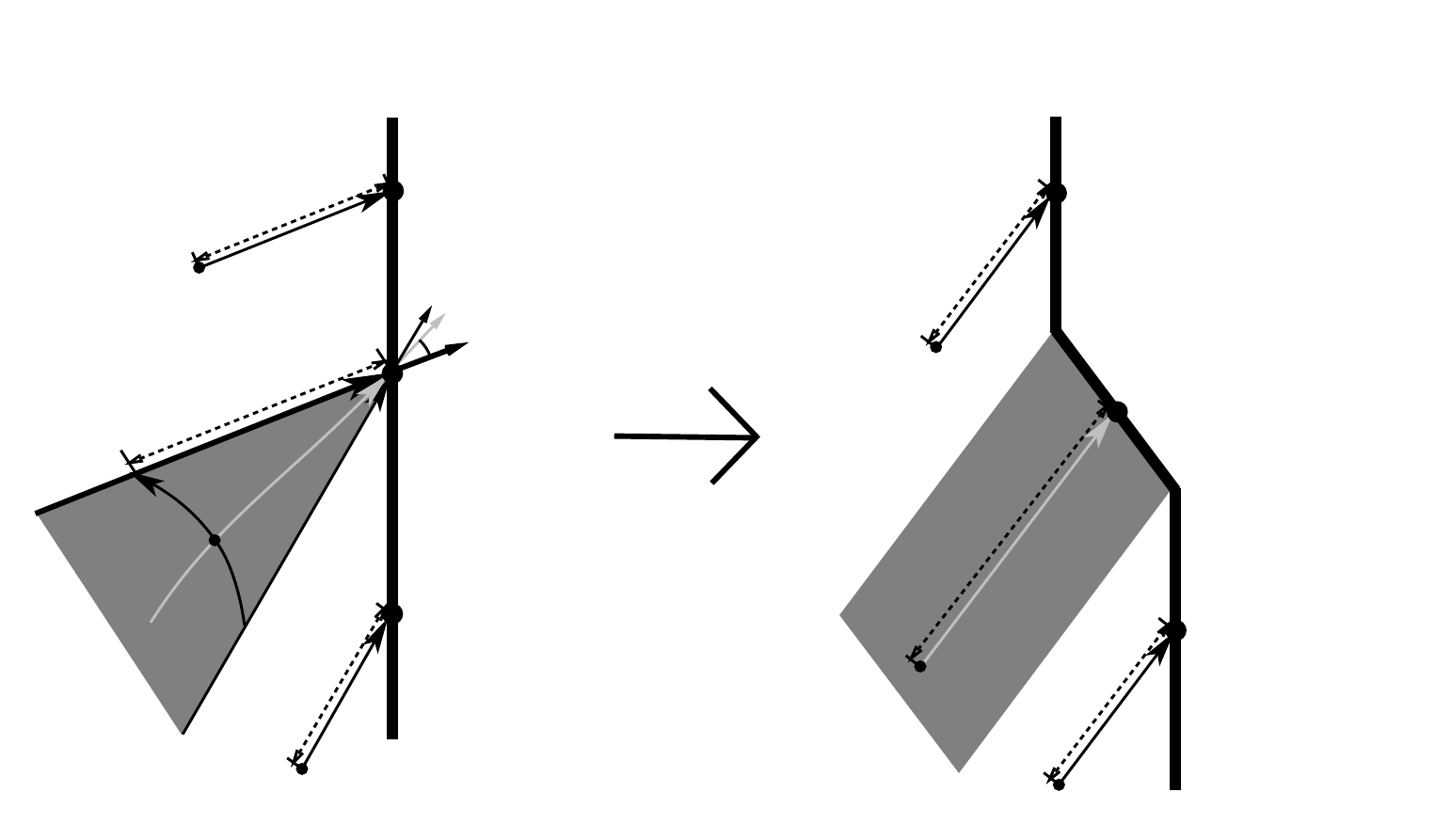}}
  \put(61.30,237.97){\fontsize{16.66}{20.00}\selectfont $V$}
  \put(135.19,155.72){\fontsize{7.07}{8.48}\selectfont $\alpha_{p_A}$}
  \put(58.41,131.86){\fontsize{5.30}{6.36}\selectfont $r_{p_A}$}
  \put(70.28,87.30){\fontsize{7.07}{8.48}\selectfont $p_A$}
  \put(56.17,169.71){\fontsize{7.07}{8.48}\selectfont $p_B$}
  \put(125.45,200.51){\fontsize{7.07}{8.48}\selectfont $(t_{p_B},0)$}
  \put(76.18,193.20){\fontsize{7.07}{8.48}\selectfont $r_{p_B}$}
  \put(125.08,69.28){\fontsize{7.07}{8.48}\selectfont $(t_{p_C},0)$}
  \put(89.11,15.62){\fontsize{7.07}{8.48}\selectfont $p_C$}
  \put(86.90,46.22){\fontsize{7.07}{8.48}\selectfont $r_{p_C}$}
  \put(327.92,200.51){\fontsize{7.07}{8.48}\selectfont $(t_{p_B},0)$}
  \put(293.24,185.14){\fontsize{7.07}{8.48}\selectfont $r_{p_B}$}
  \put(342.60,134.55){\fontsize{7.07}{8.48}\selectfont $(-\alpha_{p_A},\alpha_{p_A})$}
  \put(296.65,98.11){\fontsize{7.07}{8.48}\selectfont $r_{p_A}$}
  \put(327.13,46.22){\fontsize{7.07}{8.48}\selectfont $r_{p_C}$}
  \put(365.79,69.97){\fontsize{7.07}{8.48}\selectfont $(t_{p_C}-a,a)$}
  \put(266.02,237.22){\fontsize{16.66}{20.00}\selectfont $V'$}
  \put(275.40,142.90){\fontsize{7.07}{8.48}\selectfont $f(p_B)$}
  \put(272.64,46.28){\fontsize{7.07}{8.48}\selectfont $f(p_A)$}
  \put(316.12,9.70){\fontsize{7.07}{8.48}\selectfont $f(p_C)$}
  \put(40.43,59.65){\fontsize{7.07}{8.48}\selectfont $\gamma_{p_A}$}
  \put(44.47,98.27){\fontsize{7.07}{8.48}\selectfont $\sigma_{p_A}$}
  \put(123.41,137.25){\fontsize{7.07}{8.48}\selectfont $(0,0)$}
  \put(325.11,156.55){\fontsize{7.07}{8.48}\selectfont $(0,0)$}
  \put(199.43,140.04){\fontsize{11.76}{14.11}\selectfont $f$}
  \put(7.67,166.94){\fontsize{11.76}{14.11}\selectfont (B)}
  \put(19.91,87.13){\fontsize{11.76}{14.11}\selectfont (A)}
  \put(55.81,8.20){\fontsize{11.76}{14.11}\selectfont (C)}
  \put(262.49,174.96){\fontsize{11.76}{14.11}\selectfont (B')}
  \put(264.12,66.33){\fontsize{11.76}{14.11}\selectfont (A')}
  \put(300.11,17.29){\fontsize{11.76}{14.11}\selectfont (C')}
  \put(361.57,107.19){\fontsize{7.07}{8.48}\selectfont $(-a,a)$}
  \put(84.74,178.49){\fontsize{7.07}{8.48}\selectfont $\gamma_{p_B}$}
  \put(102.79,31.90){\fontsize{7.07}{8.48}\selectfont $\gamma_{p_C}$}
  \put(5.67,108.91){\fontsize{7.07}{8.48}\selectfont $\gamma_{-1/2}$}
  \put(59.79,32.37){\fontsize{7.07}{8.48}\selectfont $\gamma_{-1}$}
  \end{picture}%
\else
  \setlength{\unitlength}{1bp}%
  \begin{picture}(443.32, 256.66)(0,0)
  \put(0,0){\includegraphics{aplicacionconforme}}
  \put(61.30,237.97){\fontsize{16.66}{20.00}\selectfont $V$}
  \put(135.19,155.72){\fontsize{7.07}{8.48}\selectfont $\alpha_{p_A}$}
  \put(58.41,131.86){\fontsize{5.30}{6.36}\selectfont $r_{p_A}$}
  \put(70.28,87.30){\fontsize{7.07}{8.48}\selectfont $p_A$}
  \put(56.17,169.71){\fontsize{7.07}{8.48}\selectfont $p_B$}
  \put(125.45,200.51){\fontsize{7.07}{8.48}\selectfont $(t_{p_B},0)$}
  \put(76.18,193.20){\fontsize{7.07}{8.48}\selectfont $r_{p_B}$}
  \put(125.08,69.28){\fontsize{7.07}{8.48}\selectfont $(t_{p_C},0)$}
  \put(89.11,15.62){\fontsize{7.07}{8.48}\selectfont $p_C$}
  \put(86.90,46.22){\fontsize{7.07}{8.48}\selectfont $r_{p_C}$}
  \put(327.92,200.51){\fontsize{7.07}{8.48}\selectfont $(t_{p_B},0)$}
  \put(293.24,185.14){\fontsize{7.07}{8.48}\selectfont $r_{p_B}$}
  \put(342.60,134.55){\fontsize{7.07}{8.48}\selectfont $(-\alpha_{p_A},\alpha_{p_A})$}
  \put(296.65,98.11){\fontsize{7.07}{8.48}\selectfont $r_{p_A}$}
  \put(327.13,46.22){\fontsize{7.07}{8.48}\selectfont $r_{p_C}$}
  \put(365.79,69.97){\fontsize{7.07}{8.48}\selectfont $(t_{p_C}-a,a)$}
  \put(266.02,237.22){\fontsize{16.66}{20.00}\selectfont $V'$}
  \put(275.40,142.90){\fontsize{7.07}{8.48}\selectfont $f(p_B)$}
  \put(272.64,46.28){\fontsize{7.07}{8.48}\selectfont $f(p_A)$}
  \put(316.12,9.70){\fontsize{7.07}{8.48}\selectfont $f(p_C)$}
  \put(40.43,59.65){\fontsize{7.07}{8.48}\selectfont $\gamma_{p_A}$}
  \put(44.47,98.27){\fontsize{7.07}{8.48}\selectfont $\sigma_{p_A}$}
  \put(123.41,137.25){\fontsize{7.07}{8.48}\selectfont $(0,0)$}
  \put(325.11,156.55){\fontsize{7.07}{8.48}\selectfont $(0,0)$}
  \put(199.43,140.04){\fontsize{11.76}{14.11}\selectfont $f$}
  \put(7.67,166.94){\fontsize{11.76}{14.11}\selectfont (B)}
  \put(19.91,87.13){\fontsize{11.76}{14.11}\selectfont (A)}
  \put(55.81,8.20){\fontsize{11.76}{14.11}\selectfont (C)}
  \put(262.49,174.96){\fontsize{11.76}{14.11}\selectfont (B')}
  \put(264.12,66.33){\fontsize{11.76}{14.11}\selectfont (A')}
  \put(300.11,17.29){\fontsize{11.76}{14.11}\selectfont (C')}
  \put(361.57,107.19){\fontsize{7.07}{8.48}\selectfont $(-a,a)$}
  \put(84.74,178.49){\fontsize{7.07}{8.48}\selectfont $\gamma_{p_B}$}
  \put(102.79,31.90){\fontsize{7.07}{8.48}\selectfont $\gamma_{p_C}$}
  \put(5.67,108.91){\fontsize{7.07}{8.48}\selectfont $\gamma_{-1/2}$}
  \put(59.79,32.37){\fontsize{7.07}{8.48}\selectfont $\gamma_{-1}$}
  \end{picture}%
\fi \caption{\label{fig3}This figure represents a conformal map
between the spacetime $(V,g)$ (at the left) and an open region
$V'$ of Minkowski spacetime (at the right).}
\end{figure}

Our example can be understood more clearly as the spacetime
$(V,g)$ is conformal (thus, isocausal) to the following open
region of Minkowski spacetime:
\[
V'={\mathbb L}^{2}\setminus (\{x\geq a\}\cup \{t+x\geq 0,\;
0\leq x\leq a\}),
\]
where $a=(\pi/2) -\arctan(1/2)$. A conformal map
$f:(V,g)\rightarrow (V',g_{0})$ is represented in Figure
\ref{fig3}, and can be described as follows.

\smallskip

The spacetime $(V,g)$ is divided in three regions: (A) the wedge
(i.e., the region between $\gamma_{-1}$ and $\gamma_{-1/2}$), (B) the
region above the wedge (above $\gamma_{-1/2}$), and (C) the region
below the wedge (below $\gamma_{-1}$). Accordingly, the spacetime
$(V',g_{0})$ is also divided in three regions: (A') $\{(t,x)\in
V': -2a\leq t-x\leq 0\}$, (B') $\{(t,x)\in V': t-x\geq 0\}$ and
(C') $\{(t,x)\in V': t-x\leq -2a\}$. Given a point $p_{A}$ of the
region (A), there exist two lightlike geodesics
$\gamma_{_{p_{A}}}$, $\sigma_{p_{A}}$ passing through it, which
are integral curves of the lightlike vector fields
$X(t,x)=(\sqrt{\beta{(t/x)}},1)$,
$Y(t,x)=(\sqrt{\beta{(t/x)}},-1)$, resp. These curves determine
the parameters $r_{p_{A}}$ (the natural Euclidean distance from
$\sigma_{p_A}\cap \gamma_{-1/2}$ to the origin) and
$\alpha_{p_{A}}$ (the Euclidean angle between the velocities of
$\gamma_{-1/2}$ and $\gamma_{p_{A}}$ at the origin), as indicated
in the figure. Then, the image $f(p_{A})$ is defined as the point
in the region $(A')$ which lies in the line $t-x=-2\alpha_{p_A}$
at the natural Euclidean distance $r_{p_{A}}$ from
$(-\alpha_{p_A}, \alpha_{p_A})$. Next, given a point $p_{B}$ in
region $(B)$, it is clearly determined by the parameters
$t_{p_{B}}$ (where $(t_{p_{B}},0)$ is the future endpoint of the
integral curve of the lightlike vector field $X$ through $p_B$)
and $r_{p_{B}}$ (Euclidean distance to this endpoint from $p_B$).
Then, the image $f(p_{B})$ is defined as the point in region
$(B')$ determined by the analogous parameters for an integral
curve of $\partial_t+\partial_x$, as indicated in the figure.
Finally, for any $p_{C}$ belonging to region $(C)$ we proceed
similarly to obtain parameters $r_{p_{C}}$, $t_{p_{C}}$, and
define $f(p_{C})$ in the region $(C')$ of $(V',g_{0})$ as the
point determined by
 $t_{p_{C}}-a$ (which selects an integral curve of
 $\partial_t+\partial_x$)
 and $r_{p_{C}}$ (which selects a point in this curve).

\smallskip

Recall that this map $f$ is obviously continuous and piecewise
smooth. Its  conformal character is ensured as it clearly maps
lightlike curves in $(V,g)$ into lightlike curves in $(V',g_{0})$.

\section*{Acknowledgments}
The authors would like to acknowledge  Prof. Senovilla for very
useful discussions and comments. The second-named author also
thanks the kind hospitality of Department of Theoretical Physics
and History of Science, University of Basque Country, during a
research stay associated to this work.

\smallskip

All the authors are partially supported by the research projects
with FEDER funds MTM2010-18099 (Spanish MICINN) and P09-FQM-4496
(Regional J. Andaluc\'{\i}a). Also, the second-named author is
supported by Spanish MEC Grant AP2006-02237.


\begin{thebibliography}{99}

\bibitem{AF} V. Ala\~na, J.L. Flores,
The causal boundary of product spacetimes, {\em Gen. Rel.
Gravitation} {\bf 39} (2007), no. 10, 1697--1718.

\bibitem{AH} A. Ashtekar, R.O. Hansen, A unified treatment of null and spatial infinity in general
relativity. I. Universal structure, asymptotic symmetries, and
conserved quantities at spatial infinity, {\em J. Math. Phys.}
{\bf  19} (1978), no. 7, 1542--1566.

\bibitem{BEE} J.K. Beem, P.E. Ehrlich, K.L. Easley,
{\em Global Lorentzian geometry}, Monographs Textbooks Pure Appl.
Math. {\bf 202} (Dekker Inc., New York, 1996).


%
%
%
%
%
%

\bibitem{Chr} P.T. Chrusciel, Conformal Boundary Extensions of Lorentzian
Manifolds, {\it J. Diff. Geom.} {\bf 84} (2010) 19-44.


\bibitem{F} J.L. Flores, The Causal Boundary of spacetimes revisited, {\it Commun. Math. Phys.} {\bf 276} (2007) 611--643.

\bibitem{FH} J.L. Flores, S.G. Harris, Topology of the
causal boundary for standard static spacetimes, {\em Class. Quant.
Grav.} {\bf 24} (2007), no. 5, 1211--1260.

\bibitem{FHSconf}  J.L. Flores, J. Herrera, M. S\'anchez, On the final definition of the causal boundary and its relation with the conformal
boundary, {\em Adv. Theor. Math. Phys.} {\bf 15} (2011) 991-1058.

\bibitem{FHSst}  J.L. Flores, J. Herrera, M. S\'anchez, Gromov, Cauchy and causal boundaries for Riemannian, Finslerian and
Lorentzian manifolds (2010), to appear in {\em Memoirs of the
AMS}. Available at arXiv:1011.1154.


%

\bibitem{JHEP}  J.L. Flores, M. S\'anchez, The causal boundary of wave-type spacetimes
{\em J. High Energy Physics}, 03 (2008) 036.

\bibitem{GSan} A. Garc\'{i}a-Parrado, M. S\'anchez,
Further properties of causal relationship: causal structure
stability, new criteria for isocausality and counterexamples
Class. Quantum Grav. 22 (2005)  4589-4619.


\bibitem{GS} A. Garc\'{i}a-Parrado, J.M.M. Senovilla, Causal relationship: A new
tool for the causal characterization of Lorentzian manifolds, {\em
Class. Quantum Grav.} {\bf 20} (2003) 625-64.


\bibitem{GSa} A. Garc\'{i}a-Parrado, J.M.M. Senovilla, Causal
symmetries, Class.Quant.Grav. 20 (2003) L139.

\bibitem{GSb} A. Garc\'{i}a-Parrado, J.M.M. Senovilla, General study and basic properties of causal symmetries, Class.Quant.Grav. 21 (2004) 661-696.

\bibitem{GS2} A. Garc\'{i}a-Parrado, J.M.M. Senovilla, Causal
structures and causal boundaries, Class.Quant.Grav. 22 (2005)
R1-R84.


\bibitem{GKP}
R.P. Geroch, E.H. Kronheimer and R. Penrose, Ideal points in
spacetime, {\it Proc. Roy. Soc. Lond. A} {\bf 237} (1972) 545--67.


\bibitem{H1} S.G. Harris,
Universality of the future chronological boundary, {\em J. Math.
Phys.} {\bf 39} (1998), no. 10, 5427--5445.

\bibitem{H2} S.G. Harris, Topology of the future chronological boundary: universality for spacelike boundaries,
{\em Classical Quantum Gravity} {\bf 17} (2000), no. 3, 551--603.

\bibitem{H} S.G. Harris, Causal boundary for standard static spacetimes, {\em Nonlinear Anal.} {\bf 47} (2001),
2971-81.




%
%
%

\bibitem{MR1} D. Marolf, S. Ross, Plane Waves: To infinity and
beyond! {\it Class. Quant. Grav.} {\bf 19} (2002) 6289--6302.

\bibitem{MR} D. Marolf, S.R. Ross, A new recipe for causal
completions, {\it Class. Quant. Grav.} {\bf 20} (2003) 4085--4117.

\bibitem{MS} E. Minguzzi, M. S\'{a}nchez,  The causal hierarchy of
spacetimes, in {\em Recent developments in pseudo-Riemannian Geometry} (2008) 359--418. 
ESI Lect. in Math. Phys., European Mathematical Society Publishing
House. (Available at gr-qc/0609119).

%

\bibitem{O} B. O'Neill, {\em Semi-Riemannian Geometry with applications to Relativity}, Academic Press, INC, 1983.

%
%


\bibitem{S} M. S\'anchez, Causal boundaries and holography on wave
type spacetimes, {\em Nonlinear Anal.}, {\bf 71} (2009),
e1744-e1764.


%
%
%
%
%




\end{thebibliography}
\end{document}